# How E-Mental Health Services Benefit University Students with ADHD: A literature Review

**Bader Binhadyan**
School of Business IT & Logistics
RMIT University Australia
Email: ben.hadyan@rmit.edu.au

**Bill Davey**
School of Business IT & Logistics
RMIT University Australia
Melbourne, Australia.
Email: bill.davey@rmit.edu.au

**Nilmini Wickramasinghe**
Epworth HealthCare and Deakin University
Melbourne, Australia
Email: nilmini.wickramasinghe@epworth.org.au

## Abstract

E-mental health is an area within e-health in which the key role of IS/IT has not been well established. Both clinicians and scholars are uncertain as to the role of IS/IT and its potential benefits. This research review is introduced to assist in understanding the enabling role in e-mental health and it focused on one area of mental health, Attention Deficit Hyperactivity Disorder (ADHD) in university students. ADHD is estimated to affect approximately 6% of university students by negatively impacting students' academic performance and social life; hence, fewer of these students complete their degrees as compared to their peers. This review outlines the potentials of using IS/IT in the treatment of ADHD in university students. It also serves to greatly contribute to e-mental health development and thereby to help to uncover further possibilities of IS/IT to support broad areas within mental health disorders and services.

## 1. Introduction

Using Information Systems and Technology (IS/IT) in healthcare, in a form of e-health, can facilitate and overcome issues in many areas of healthcare(Wickramasinghe et al. 2005). One of the areas of healthcare that has started to take advantage of IS/IT capabilities is mental health. E-mental health (eMH) has the potential to deliver easy access and cost effective treatments, and early interventions and support to people with different mental disorders (Lal and Adair 2014). Current mental health service can be enhanced by using 'sensible technologies' such as Short Message Service (SMS), Email, website, Chat/ Instant Messaging (IM) tools, social media, video/audio over the Internet, as well as smartphones and tablets (Whittaker et al. 2012). Although there appears to be great potential for eMH in the context of young adults with mental disorders (Burns et al. 2010; Webb et al. 2008), the majority of eMH programs are mostly targeting illnesses related to depression, anxiety and suicidal thoughts (Orman et al. 2014). Hence, there appears to be great potential in using eMH in the contexts so far not explored, including ADHD in university students.

ADHD a chronic mental disorder that affects roughly 6% of the university student population and presents with symptoms of hyperactivity, inattention, daydreaming, and impulsiveness (McGillivray and Baker 2009). Studies have indicated that there is a relationship between ADHD symptoms and university students' academic underperformance (Pope 2010) with  fewer of such students completing their degree in comparison with non-ADHD students. There are issues that university students with ADHD face during their studies such as finding university environment and social life overwhelming or having deficits in organisation and study skills as well as in executive functions in the brain such as planning and working memory (Gropper and Tannock 2009). Executive function has been found to have a direct relationship with the academic performance of students.

This paper is to review current research on what is already known about the use of IS/IT in a form of eMH. It also addresses the issues related to university students with ADHD and outlines the potential use of eMH in the treatment of ADHD in university students. The research question which the article aims to answer is "*what are the main streams in current research that address the trend of e-mental health and the issues associated with ADHD?*"

Previous literature reviews outlined the benefits of using of eMH to improve services and treatment delivery for mental disorders such as depression, anxiety, suicidal thoughts (Karasouli and Adams 2014; Lal and Adair 2014; Musiat and Tarrier 2014); however, there is no similar work that has been conducted on literature around the use of eMH in the treatment of university students with ADHD.

## 2. Research Methodology and Design

The research methodology used in this paper to explore the streams of research that address the role of IS/IT in mental health and the issues associated with ADHD in university students is systematic literature review (Petticrew and Roberts 2006). The research used the available research engines at the university library. The databases that were used include ACM Digital library, ScienceDirect, IEEE Explore, and SpringerLink, Academic Research Library (ProQuest), and Australian Standards. Google Scholar was used to complement this process. Since the use of IS/IT in mental health is quite recent, there was no date limit applied to any research in the broader area. However, the publications related to the ADHD part of this research were limited to publications after 2004.

The keywords and their combinations that were employed in this research were, *e-mental health, IS/IT, Internet, technology, mobile, mental health, ADHD, treatments, college,* and *university*. Title, keywords and abstract were the searched areas. Conferences papers and journal articles reporting research addressing the trend of eMH and the issues associated with ADHD were identified. Initially, over 260 articles were indexed. However, this literature review is limited to journal articles that have been published that were not based on medication experiments or biology studies. Medication was seen to be potentially monitored by IS/IT but not to be an eMH treatment. Because this research focuses on adults with ADHD, or university students with ADHD, publications that were based on children under the age of 16 were also eliminated. This process returned 74 articles. The process of selection is shown in Figure 1. Furthermore, articles (n=74) were later classified based on the main stream of the research, eMH, ADHD, or Mental health. Articles were focused on e-mental health (n=28), ADHD (n=47), and mental health (n=4).

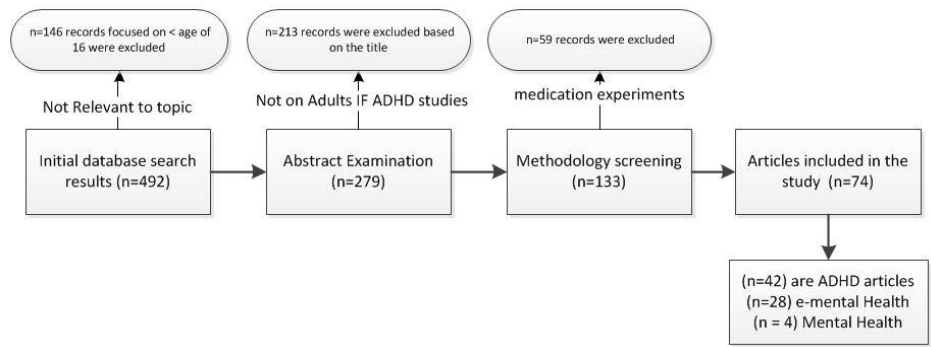

*Figure 1 Research Selection Process*

## 3. Results

### 3.1. E-Mental Health Opportunities

The literature review analysis identified validated eMH capabilities to diminish the barriers to the current mental health system. Stigma among young adults, especially university students is found to be one of the most important barriers to mental health treatments as well as financial situation, and mental health literacy (Eisenberg 2007; Gulliver et al. 2010; Hunt 2010). EMH has the ability to reduce the barriers to access current mental healthcare in a number of ways as follow:

- Provide services that can be accessed anytime and anyway with less or no cost (Booth et al. 2004).

- Reduce stigma incurred by seeing a therapist and improve mental health literacy (Burns et al. 2010; Christensen and Hickie 2010)
- Improve the therapists time and efficacy (Jorm et al. 2013; Jorm et al. 2007).
- Provide early intervention and treatment for different disorders (Christensen and Petrie 2013)

The literature showed that there are five types of eMH services in Australia (Lal and Adair 2014; Orman et al. 2014):

1. Health promotion, wellness promotion and psycho-education.
2. Prevention and early intervention
3. Crisis intervention and suicide prevention
4. Treatment
5. Recovery

Several services are in existence using available technologies to deliver eMH services or treatment. The technologies include mobile systems targeting depressions and anxiety (Ben-Zeev et al. 2012; Proudfoot 2013; Proudfoot et al. 2010; Whittaker et al. 2012), websites for suicidal thought reductions (van Spijker et al. 2012), the Internet for social phobia (Titov et al. 2008). Moreover, Psychotherapy and coaching, or education such as Cognitive Behaviour Therapy (CBT) can be delivered using one these technologies effectively to treat many disorders including ADHD (Kaltenthaler et al. 2006; Pettersson et al. 2014).

### 3.2. ADHD

Until recently, ADHD has been known as a childhood disorder but it has been found that the majority of children with ADHD will carry it into adulthood and it affects approximately 6% of the university population (DuPaul et al. 2009; Weyandt and DuPaul 2006; Weyandt and DuPaul 2008). The data analysis shows that ADHD presents as three symptoms categories, which are shown in Table 1

| Symptoms | Characteristics |
| --- | --- |
| Impulsivity | Difficulties in self-control<br>Lack of family functioning<br>Substance abuse<br>Trouble with the law |
| Hyperactivity | Restlessness, racing thoughts<br>Unreasonable talking,<br>Trying to do many things at once |
| Inattention | Inability to remain focused, concentrate, or stay on task<br>Lack of attention to detail or inability to follow directions<br>Remembering information<br>Lack of management and organisation skills |

Table 1 Characteristics of Adults with ADHD

The analysis of the literature review shows that most students with ADHD face many difficulties in adjusting to university. They might find university academic social life to be overwhelming (DuPaul et al. 2009). In addition, studies have indicated that there is a relationship between ADHD symptoms and university students' academic underperformance (Pope 2010). The literature indicates that university students with ADHD are often less functioning on a number of academic variables compared to their peers, and these variables are show in Table 2.

| Academic variables | References |
|---|---|
| Study skills - note taking, summarising, organising | (Frazier et al. 2007; Weyandt and DuPaul 2006) |
| Anxiety, depression and motivation | (Meaux et al. 2009) |
| Executive Functions deficit | (Gropper and Tannock 2009) |

Table 2 Academic variables

The analysis of the literature shows that there are common academic behaviours among university students with ADHD which are show in Table 3

| Factor | References |
|---|---|
| Higher rates of subject failure | (Meaux et al. 2009) |
| Lower Grade Point Averages | (DuPaul et al. 2009; Rabiner et al. 2008) |
| Class attendance | (DuPaul et al. 2009; Meaux et al. 2009; Weyandt and DuPaul 2006; Weyandt and DuPaul 2008) |

Table 3 Common Academic Behavioural

ADHD treatment includes multiple methods such as psychotherapy, coaching, education and medication (Ebejer et al. 2012). The tools that have been found in the literature and previous studies are listed in Table 4

| Treatment | Tools | References |
|---|---|---|
| Coaching / Education | Smartphone or SMS reminder, Email coaching and Time management | (Prevatt et al. 2011) |
| Therapy | Neurofeedback Therapy | (Arns et al. 2009; Arns et al. 2014; Lansbergen et al. 2011; Wang and Hsieh 2013) |
| | CBT/Internet-based CBT | (Pettersson et al. 2014) |

Table 4 IS/IT tools found in ADHD treatment

## 4. Discussion and Conclusion

The rapid development of sensible technologies, such as the Internet, and smart phones, has made impossible ideas to be possible and introduced methods that provide better treatments and early interventions for common mental health issues. eMH services have the ability to deliver better accessibility, reduce cost and enhanced quality, and provide a higher level of flexibility. Implementing these potentials can assist the delivery of the current treatments of ADHD especially for ADHD in university students. Some of the theme trends found in this research are 1) eMH can improve accessibility to mental health services 2) eMH provides cost effective treatment 3) eMH improve mental health literacy by providing improved delivery methods of mental health promotion 4) eMH can provide early interventions and better treatments. In the next part of this section, these themes will be outlined with regard to treatment of ADHD in university students.

### 4.1. Improved Accessibility

Research shows that improved access to information by patients or customers about their care will improve the care integration for these people and empower them (Christensen and Petrie 2013). eMH may empower the university student with ADHD in another way by extending the possibilities as to who can be involved in the delivery of mental health preventive strategies (Jorm et al. 2013). eMH also will have the ability to enable information exchange in a standardised way between mental health providers and education providers. Regarding better access to treatment, CBT method has been delivered though the internet or smart phone which can be used by students anytime and anywhere.

It has been found that CBT can be effective in treating individuals with ADHD with aims to 1) assist these individuals to overcome issues with their executive functions that are important in managing time and organisation (Baer et al. 2007; Lindstedt and Umb-Carlsson 2013) 2) focus on emotional self-regulation, impulse control, and stress management (Pettersson et al. 2014; Safren et al. 2005).

Although, eMH promises to facilitate the accessibility of mental healthcare services, people with mental health issues might find it rather overwhelming or mental health professionals who are unfamiliar with technology may limit their access to the services (Lal and Adair 2014).

### 4.2. Reduced Cost and Increased Quality

eMH has the potential to increase efficiency in mental health and reduce costs. Previous studies show the delivery of eMH is both cost-effective and cheaper in comparison to the traditional mental health delivery (Lal and Adair 2014). This is because some interventions or treatments can be delivered over the internet with no or very low cost (Orman et al. 2014).

Improving the service quality is one of the advantages of internet which can be seen in improving access by consumers, automation of data collection and processing, and making treatment independent of geographic location. eMH services have successfully been used in therapeutic situations which can be seen in enabling users to access therapy and support in real-time where and when they need them (Proudfoot 2013).

eMH will provide online sources for training and professional development of mental health professionals and delivery of psych-education for university students with ADHD (Burns et al. 2010). In education, it helps adults with ADHD to understand how such a disorder can affect multiple areas of their lives. This increased understanding of their condition may result in reducing symptoms, improving functioning and preventing negative consequences of ADHD. It has been argued that education must be the first intervention because the more educated patients are, the better their response to treatment (Safren et al. 2005). Education also can be provided as by any form of website, online video or Podcast.

### 4.3. Better Promotion Technique

Providing health promotion and psych-education programmes for people with mental health issues can be effective in improving symptoms (Riper et al. 2010). Successful adaptions and implementing of promotion programmes that provide training and information in the education sectors are related to enhancements in student's behaviours and academic performance (Dix et al. 2012), This online programme can also improve the referral process or improve seek-help especially for young adults who have not previously accessed health services (Burns et al. 2010).

### 4.4. Early Intervention and Prevention

eMH based prevention and early intervention tools have enabled university students with ADHD to better cope with their mental health. eMH services that provide different tools to deliver, such as email and/or chat, counselling and referrals can be effective in reducing distress, connecting consumers to treatment and improve motivation to act (Lal and Adair 2014). It has been found that websites can be used to as prevention tool to reduce substance abuse among students (Newton et al. 2010).

### 4.5. Conclusion

In conclusion, this review of current research shows that a wide potential exists for the use of IS/IT in Mental Healthcare. There was little literature discussing the key role of IS/IT in the treatment of ADHD in university students and other ADHD populations. This study also shows that there is little information regarding the implications of adoption of eMH by relevant professionals. This points to the need for further investigation which will contribute to the practice of using IS/IT in delivery of mental health services. One clear direction for research would be to interview psychologists to seek what possibilities are available for leveraging the role of IS/IT in ADHD treatments. This will also generate outcomes that would add to the theory regarding eMH.

## Copyright